\documentstyle[aps,twocolumn]{revtex}

\title{Effective field theories of non-equilibrium physics}

\author{Mark Burgess}

\address{Centre of Science and Technology, Oslo College, Cort Adelers gate 30, 0254 Oslo, Norway}
\def\beq{\begin{eqnarray}}
\def\eeq{\end{eqnarray}}
\def\2{\frac{1}{2}}
\def\32{\frac{\sqrt 3}{2}}
\def\i2{\frac{i}{\sqrt 2}}
\def\12{\frac{1}{\sqrt 2}}

\def\i{\hat{\bf i}}

\begin{document}

\maketitle
\begin{abstract}
Non equilibrium effective field theory is presented as an
inhomogeneous field theory, using a formulation which is analogous to
that of a gauge theory. This formulation underlines the importance of
structural aspects of non-equilibrium, effective field theories. It is
shown that, unless proper attention is paid to such structural
features, hugely different answers can be obtained for a given model.
The exactly soluble two-level atom is used as an example of both the
covariant methodology and of the conclusions.
\end{abstract}

\section{Scales and interactions}

Our conventional wisdom about isolated systems is that processes
which dominate at microscopic scales have no detailed effect on the
behaviour of a physical system at much larger scales.  This linear
viewpoint is the basis of effective field theory. Put another way it
says that, when the ratio of microscopic to macroscopic scales is very
marked, one is justified in making a continuum approximation and
blurring out the small details of the system.

This rule of thumb is not universal.  A prolific exception, to name a
single example, is the evolution of life. Biologically, very small
changes at the molecular level (genes/proteins) induce large and very
dramatic changes in the way a system develops macroscopically
(phenotype).  What starts as a few proteins bumping into one another,
ends up as a person reflecting on the nature of the development at a
physics conference.  The transition is a dramatic one and clearly
cannot be described by any simple field theory because it involves
complex interactions with time varying boundary conditions coupling a
whole hierarchy of scales. In contrast, when you look at a grown
organism, you do not see the effects of every protein movement in the
body: so there a continuum model seems to be a good idea. The
difference is that there are no longer major structural changes taking
place. So in equilibrium things are well described by a continuum
hypothesis, but during development (far from equilibrium) that it
not necessarily true.

What occurred in between was due to complicated time-dependent boundary
conditions which caused processes from a hierarchy of scales to
interact strongly. While each arbitrary part of a system follows
microscopic laws in every detail, the totality of a complex system
exceeds the sum of its parts because it involves structural
information about how to put those parts together, i.e. the boundary
conditions between neighbouring elements in a system.  Cooperation and
competition between neighbouring cells introduces huge complications
to model builders and---even in the simplest case---we probably need
an evolving effective theory, if not several, to describe even the
simplest non-equilibrium development in a reasonable fashion.

This paper is about basic structural aspect of field theories with
spacetime dependent external perturbations. The perturbations will be
treated generically and therefore they can be thought of as effective
interactions; they may be justified by any number of arguments: as the
generators of resummations in interacting field theories, as
renormalization counterterms, or as external influences whose form is
dictated by coordinate invariance and unitarity. By using a generic
Schwinger source theory, all of these possibilities are covered in a
single abstraction.  By taking a quadratic action, one considers the
simplest idealization of a system interacting with its surroundings.
Despite being the prototype for basically all interacting systems to
low perturbative order, even this simple problem is difficult and not
very well understood.  I would like to illustrate how changes in
the basic structural elements of non-equilibrium field theory can
have huge consequences for the way in which a theory behaves and leave
this both as an interesting direction for future investigation and
as a `warning as to the unwary' as to the kind of pitfalls which
one might encounter in model building away from equilibrium.

\section{Summary}

This paper contains three overlapping messages.

\begin{itemize}
\item That a proper understanding of non-equilibrium development involves
all of the subtleties and structure which gauge theories introduce.

\item That interesting similarities exist between the 
classical two level atom and Schwinger's closed time
path\cite{schwinger2} (CTP) formalism. Broadly speaking the upper and
lower levels of the two-level system correspond to creation and
annihilation in regular field theory.

\item That small changes in theoretical structure can lead to 
large changes in the behaviour of solutions.

\end{itemize}
The latter two points are exlpored using an exactly soluble model of the
two-level atom interacting with a strong radiation source.  This talk
is a summary of more detailed work contained in two
papers\cite{burgess12,burgess15}. The conventions are detailed in
reference \cite{burgess12}.

\section{Inhomogeneity}

{\em The purpose of this section is to motivate a method for the analysis
of non-equilibrium field theory, drawing structural features from
gauge theories. The action is presented in generic form, in terms of
general bi-local sources which represent the interactions which a
system may have at quadratic order.}
\vspace{0.5cm}

It might strike one as surprising that gauge theory ideas would crop
up in the study of arbitrary non-equilibrium systems, but the reason
is clear.  Disequilibrium in a physical system is associated with some
kind of inhomogeneity: either in space or in time, usually
both. Uneven perturbations and variations in physical quantities lead
to transport and relaxation.  It is thus useful to think of field
theory away from equilibrium as an inhomogeneous field theory.  Gauge
theory is about inhomogeneous, space-time dependent phases, or complex
re-scalings at arbitrary points. It is therefore also a good framework
for inhomogeneous field theory, where scale factors change at every
point. Clearly these are not the same, but they are directly analogous.

We begin then with a lightning summary of inhomogeneous field
theory, taken from ref. \cite{burgess12}.  Between any two space-time
points $x$ and $x'$ it is useful to parameterize functions in terms of
`rotated' variables:
\begin{eqnarray}
\tilde x &=& (x-x')\nonumber\\
\overline x &=& \frac{1}{2}(x+x')
\end{eqnarray}
The odd variables $\tilde x$ (the variable conjugate to the momentum)
characterize `translational invariance' while the even variables
$\overline x$ represent the opposite of this: inhomogeneity.

Many aspects of non-equilibrium field theory are usefully described in
terms of the closed time path (CTP) generating
functional\cite{schwinger2,bak,keldysh,calzetta1}. The CTP is a field
theoretical prescription for deriving expectation values of physical
quantities, given a description of the state of the field at some time
in the past. The generating functional requires an artificial
duplicity in the field, so the closed-time path action is described as
a two-component field $\phi_A$, where $A=+,-$.
\begin{equation}
S_{CTP} = \int dV_x dV_{x'} \frac{1}{2}\phi^A S_{AB} \phi^B
.\label{eq:77}
\end{equation}
The CTP field equations in the presence of sources may be found by
varying this action with respect to the $+$ and $-$ fields.  Following
a similar line of argument to Lawrie\cite{lawrie1}, a general Gaussian,
quadratic form for a closed time path action may be expressed in terms
of general sources. These sources can be thought of as external
quantities or as the renormalized shadows of higher loop contributions
due to self-interactions.  We write
\begin{eqnarray}
S_{AB}(x,x') =
\left( 
\begin{array}{cc}
\hat\alpha & \hat\beta\\
-\hat\beta & -\hat\alpha
\end{array}
\right)
\label{eq:75}
\end{eqnarray}
where the indices $A,B$ run over the $\pm$ labels of the CTP fields,
\beq
\hat\alpha &=& (-{\vcenter{\vbox{\hrule height.4pt\hbox{\vrule width.4pt height8pt\kern8pt\vrule width.4pt}\hrule height.4pt}}}+m^2)\delta(x,x')+I(x,x') \nonumber\\
\hat\beta &=& J(x,x') + K^\mu(x,x') \stackrel{\leftrightarrow}{D^K_{\mu}}'
\eeq
and a new derivative has been defined to commute with the function
$K^\mu(x,x')$:
\begin{equation}
\stackrel{x}{D_\mu^K}~ \equiv ~\stackrel{x}{\partial}_\mu + 
\frac{1}{2}\frac{\stackrel{x}{\partial}_\mu K_\nu(x,x').}{K_\nu(x,x')}
\label{eq:76}
\end{equation}
for time reversibility.
Notice the general form of the `connection' term in this
derivative. This inhomogeneous, conformal structure crops up
repeatedly in non-equilibrium development.  The currents associated
with sources $I,J,K^\mu$ are not necessarily conserved since their
behaviour is not completely specified by the action, but the action is
differentially reversible.  The sum of rows and columns in this
operator is zero, as required for unitarity and subsequent causality.
The significance of the off-diagonal terms involving $K^\mu$ can be
seen by writing out the coupling fully:
\begin{equation}
K^\mu(x,x') \cdot \left( \phi_1 D^K_\mu \phi_2 - \phi_2 D^K_\mu \phi_1\right).
\label{eq:gamma}
\end{equation}
The term in parentheses has the form of a current between components
$\phi_1$ (the the forward moving field) and $\phi_2$ (the backward
moving field).  When these two are in equilibrium there will be no
dissipation to the external reservoir and these off-diagonal terms
will vanish. This indicates that these off-diagonal components (which
are related to off-diagonal density matrix elements, as noted earlier)
can be understood as the mediators of a detailed balance condition for
the field.  When the term is non-vanishing, it represents a current
flowing in one particular direction, pointing out the arrow of time
for either positive or negative frequencies.  The current is a
`canonical current' and is clearly related to the fundamental
commutator for the scalar field in the limit $+\rightarrow -$.

The system can be analyzed by looking for the Green functions
associated with this system. These can all be expressed in terms of the
Wightman functions $G^{(\pm)}(x,x')$ using the relations
\begin{equation}
G^{(+)}(x,x') = \left[G^{(-)}(x,x') \right]^* = - G^{(-)}(x',x).
\label{eom26}
\end{equation}
The Wightman functions are the sum of all positive or negative
energy solutions, satisfying the closed time path field equations,
found by varying the action above. Thus they are the embodiment
of the dispersion relation between $\bf k$ and $\omega = k^0$.
\begin{eqnarray}
\tilde G(x,x') = G^{(+)}(x,x') + G^{(-)}(x,x')\nonumber\\
\overline G(x,x') = G^{(+)}(x,x') - G^{(-)}(x,x').\label{commG}
\end{eqnarray}
$\overline G(x,x')$ is the sum of all solutions to the free field
equations and, in quantum field theory, becomes the so-called {\em
anti-commutator} function. The symmetric and anti-symmetric
combinations satisfy the identities
\begin{equation}
\stackrel{x'}{\partial_t} \overline G(x,x')\Bigg|_{t=t'} = 0,
\end{equation}
and
\begin{equation}
\stackrel{x'}{\partial_t} \tilde G(x,x')\Bigg|_{t=t'} = \delta({\bf x},{\bf x}').\label{commrelG}
\end{equation}
The latter is the classical dynamical equivalent of the fundamental
commutation relations in the quantum theory of fields.  Other Green
functions may be constructed from these to model the processes of
emission, absorption and fluctuation respectively:
\begin{eqnarray}
G_r(x,x') &=& -\theta(t,t')\tilde G(x,x')\nonumber\\
G_a(x,x') &=& \theta(t',t)\tilde G(x,x')\nonumber\\
G_F(x,x') &=& -\theta(t,t')G^{(+)}(x,x') + \theta(t',t)G^{(-)}(x,x').\label{eom23}
\end{eqnarray}
It may be verified that, since $G^{(+)}(k,\overline x)$ depends only
on the average coordinate, the commutation relations are preserved
(see equation (\ref{commrelG})) even with a time-dependent action.
The general solution for the positive frequency Wightman function may
be written
\begin{equation}
G^{(+)}(x,x') = -2\pi i\int \frac{d^{n-1}k}{(2\pi)^{n-1}}e^{2ik_\mu\tilde x^\mu}
\frac{(1+f(k_0,\overline x))}{2|\omega|}
\label{eq:88}
\end{equation}
where $f(k_0,\overline x)$ is an unspecified function of its arguments
and it is understood that $k_0=|\omega|$ (this describes the dispersion
of the plane wave basis). The ratio of Wightman functions describes
the ratio of emission and absorption of a coupled reservoir (the
sources). As observed by Schwinger\cite{martin1,schwinger5}, all fluctuations
may be thought of as arising from generalized sources via the Green
functions of the system. Thus a source theory is an effective description
of an arbitrary statistical system.

In an isolated system in thermal equilibrium, we expect the number of
fluctuations excited from the heat bath to be distributed according to
a Boltzmann probability factor\cite{reif1}.
\index{Boltzmann factor}
\beq
\frac{\rm Emission}{\rm Absorption} = \frac{-G^{(+)}(\omega)}{G^{(-)}(\omega)}
= e^{\hbar\beta|\omega|}\label{KMS1}.
\eeq
$\hbar\omega$ is the energy of the mode with frequency $\omega$.  This
is a classical understanding of the well-known Kubo-Martin-Schwinger
relation\cite{kubo1,martin1} from quantum field theory. In the usual
derivation, one makes use of the quantum mechanical time-evolution
operator and the cyclic property of the trace to derive this relation
for a thermal equilibrium. The argument given here is identical to
Einstein's argument for stimulated and spontaneous emission in a
statistical two state system, and the derivation of the well-known $A$
and $B$ coefficients. It can be interpreted as the relative occupation
numbers of particles with energy $\hbar\omega$. This is a first hint
that there might be a connection between heat-bath physics and the two
level system.

Finally, it is useful to define quantities of the form
\begin{eqnarray}
F_\mu &=& \frac{\partial_\mu f}{f} = \frac{1}{2}\overline \partial_\mu \ln(f)\label{eq:89}\\
\Omega_\mu &=& \frac{\partial_\mu\omega}{\omega} =  \frac{1}{2}\overline\partial_\mu\ln|\omega(\overline x)|.\label{eq:90}
\end{eqnarray}
which occur repeatedly in the field equations and dispersion relations
for the system and characterize the average rate of development of the
system.  Note the similarity in form to the connection term in the
derivative of eqn. (\ref{eq:76}).

\section{Inhomogeneous scaling and gauge formulation}

{\em General covariance in an inhomogeneous environment requires
one to acknowledge the existence of a connection which transforms,
by analogy with a gauge theory.}
\vspace{0.5cm}

Inhomogeneous field theory can be presented in a natural form by
introducing a `covariant derivative' $D_\mu$ which commutes with the
average development of the field and is physical in the sense of being
Hermitian in the presence of the sources.  This description parallels
the structure of a gauge theory (in momentum space) with a complex
charge. One may also speak of a generalized chemical potential or of a
special case of quantum field theory in curved spacetime (see the
local momentum space expansion approach of ref. \cite{calzetta1} as
well as the paper by A.G. Nicola in these proceedings). Consider the
derivative
\begin{equation}
D_\mu = \partial_\mu - a_\mu
\label{eq:200}
\end{equation}
and its square
\begin{equation}
D^2 = {\vcenter{\vbox{\hrule height.4pt\hbox{\vrule width.4pt height8pt\kern8pt\vrule width.4pt}\hrule height.4pt}}} - \partial^\mu a_\mu - 2a^\mu\partial_\mu + a^\mu a_\mu.
\label{eq:201}
\end{equation}
Derivatives occur in the field equations and in the dispersion relation
for the field and they are thus central to the dynamics of the field
and the response (Green) functions. As with a gauge theory, the effect of
derviatives on spacetime dependent factors may be accounted for
in a number of equivalent ways, by redefinitions of the field. In a gauge
theory, we call this a gauge transformation and we usually demand that
the theory be covariant, if not invariant under such transformations.
In a non-equilibrium field theory, we require only covariance, since
it is normal to deal with partial systems in which conserved currents
are not completely visible and thus invariance need not be
manifest.

In order to solve the closed time path field equations, it is 
useful to solve the dispersion relation, giving the physical
spectrum of quasi-particles in the system. In the Keldysh diagrammatic
expansion of Schwinger's closed time path generating functional,
one expands around free particle solutions. By starting with a
quasi-particle basis here we can immediately take advantage of
resummations and renormalizations which follow from the unitary
structure (specifically two-particle irreducible or daisy diagrams).
It also allows one to track changes in the statistical distribution
through the complex dispersion relation, instead of using 
real Vlasov equations coupled to real equations of motion.
Without any approximation, it is straightforward to show that, in the
general inhomogeneous case,
\begin{eqnarray}
(-{\vcenter{\vbox{\hrule height.4pt\hbox{\vrule width.4pt height8pt\kern8pt\vrule width.4pt}\hrule height.4pt}}}+m^2)G^{(+)}(x,x') = &-&2\pi i \int \frac{d^{n-1}k}{(2\pi)^{n-1}}
\frac{(1+f)}{2|\omega|}
e^{ik(x-x')}\nonumber\\
\Big\lbrack - (ik_\mu + F_\mu-\Omega_\mu)^2 &-& \partial^\mu(ik_\mu + F_\mu -\Omega_\mu)\Big\rbrack = 0.
\label{eq:202}
\end{eqnarray}
It is then natural to make the identification
\begin{eqnarray}
a_\mu &=& F_\mu - \Omega_\mu +\overline K_\mu \nonumber\\
      &=& -\partial_\mu S_E(k) +\overline K_\mu
\label{eq:203}.
\end{eqnarray}
This expression shows that the connection embodies the effect of
changing statistical distributions and quasi-particle energies
($\omega$ is solved in terms of the sources through the dispersion
relation). It also shows that the source $K^\mu$ plays essentially the
same role as these effects and it thus capable of `resumming' them.
Furthermore, the field $a_\mu$ is related to the rate of increase of
the entropy $S_E$.  In terms of the covariant derivative, one now has:
\begin{eqnarray}
(-D^2+m^2)G^{(+)}(x,x') &=& -2\pi i\int \frac{d^{n-1}k}{(2\pi)^{n-1}}\frac{(1+f)}{2|\omega|}\nonumber\\
\lbrace - (ik_\mu -\overline K_\mu)^2) &-& \partial^\mu(ik_\mu-\overline K_\mu)\rbrace
.
\label{eq:204}
\end{eqnarray}
The differential equation satisfied by $G^{(+)}(x,x')$ is thus
\begin{eqnarray}
\Big[ -D^2+m^2 &+&\overline K^2(k,\overline x) +\overline I(k,\overline x) 
- \tilde J(k)\nonumber\\
 &+& \frac{i}{2}(\overline\partial_\mu\overline I)(T^\mu-v_g^\mu/\omega) \Big]_k G^{(+)}(k,\overline x) = 0
\label{eq:205}
\end{eqnarray}
where the appearance of the subscript $k$ to the bracket serves to
remind that the equation exists under the momentum integral.  The
positive frequency field may be `gauge transformed' using the
integrating factor (Wilson line)
\begin{eqnarray}
\phi(k) &\rightarrow & \phi(k) e^{\int a_\mu d\overline x^\mu}
\label{eq:208}
\end{eqnarray}
This shows the explicit decay (amplification) of the $k$-th field
mode. This transformation also has a nice physical interpretation in
terms of the entropy of the models, defined above. The Wilson line is
the negative exponential of the entropy, showing how the field decays as
the energy of a mode becomes unavailable for doing work, i.e. as its
entropy rises.

One should not confuse these transformations with similarity
transformations on the closed time path action. Because of unitarity,
the plus and minus components of closed time path fields satisfy a
global $O(1,1)$ symmetry, which allows a certain freedom in the way
one chooses to set up the solution of the system. One can choose, for
instance, to work with Feynman Green functions and Wightman functions,
or with advanced and retarded functions, or with general mixtures of
these. The only constraint imposed by unitarity is that the sum of
rows and columns in the action (i.e.  in the argument of the
exponential in the generating functional) remains zero when plus and
minus labels are removed.  The transformations considered here are
simply field rescalings. This need not even be a symmetry of the
system. Symmetries do not have a monopoly on covariance, indeed
covariance is especially important for changes which do not leave the
system invariant. All reparameterizations of a theory, be they gauge
transformations or field redefinitions demand this.  The reason for
the similarity in form between a gauge theory and a theory of field
rescalings is that both are linked through the conformal group. This
also makes the connection to the curved spacetime approach already
referred to. General spacetime metrics are not of interest, but
conformal rescalings are.  Gauge theories bear the structure of the
conformal group, not the Lorentz group and time dependent
perturbations and changes of variable are also connected with
inhomogeneous rescalings of the conformal group.  One will not
normally see a conformal symmetry in the original action because the
effective field theories we are discussing are incomplete: they
describe partial systems, in which we ignore heat baths and
external influences etc. One can, of course, argue that there are
reasons to generalize desriptions of non-equiilrbium field theory such
that they are fully covariant with respect to such
transformations. That is a central observation of this paper.

It is, of course, natural to suppose that changes of variable,
i.e. changes of perspective will lead to transformations analogous to
gauge transformations. After all, we are perturbing a system with
sources which vary in space and time. Methods of solution which rely
on diagonalization of the action will also involve transformations
which depend on space and time. All such transformations demand
covariant derivatives and transforming auxiliary fields.  The
amplification of modes makes this a recipe for a kind of space-time
dependent renormalization group.  Some authors have suggested making
coupling constants run with time, but there are canonical restrictions
associated with making coupling constants depend on space-time
coordinates\cite{burgess16}. The idea of completion by general
covariance is also reminiscent of the Vilkovisky-DeWitt effective
action\cite{vilkovisky}; perhaps this also has an interpretation in non-equiibrium
physics.

An important question associated with this new gauge-like formulation
is therefore: should we introduce `gauge' fields from the start? If so, what
initial value should they have? To demonstrate the importance of this
issue, as well as to provide an example of the approach,
I would now like to turn to a directly analogous problem which is
closely related to hard experimental data for the micro-maser.

\section{Two state system}

{\em The purpose of this section is to explore the meaning
of the `connection' in the context of an exactly
soluble model, and to show that its presence has intriguing
effects on the nature of the theory. The suggestion is that
a more careful understanding of the role of this connection
and its conformal structure could lead to improved calculational
schemes in more general cases.}
\vspace{0.5cm}

The two level atoms is not, in itself, important to the message of
this paper. Any exactly soluble model would suffice, though such
benisons are hard to come by. It does have several advantages however.
In particular, the two level structure has a strikingly similar structure
to the plus, minus labels of the closed time path generating
functional.

The Jaynes-Cummings model\cite{jaynes1} is the archetypal model for
the interaction between a two-level atom and monochromatic light.  It
derives from a {\em simplification} of the phenomenological two-level system
described by the action
\beq
S &=& \int dV_x \Big[ 
-\frac{\hbar^2}{2m}(\partial^i\psi_A)^*(\partial_i\psi_A) \nonumber\\
&-& \psi^*_A V_{AB}(t)\psi_B+\frac{i\hbar}{2}(\psi^* D_t \psi - (D_t\psi)^*\psi)
\Big]\label{eq1}
\eeq
where $A,B=1,2$ characterizes the two levels, $i\hbar D_t =
i\hbar\partial_t + i\Gamma(t)$ in matrix notation, and $\Gamma_{AB}$
is an off-diagonal anti-symmetrical matrix.  At frequencies which are
small compared to the light-size of atoms, an atom may be considered
electrically neutral: the internal distribution of charge is considered
to be irrelevant. In this approximation the leading interaction is a
resonant dipole transition.  The connection $\Gamma_{AB}$ plays an
analogous role to the electromagnetic vector potential in
electrodynamics, but it possesses no dynamics of its own. Rather it
works as a constraint variable, or auxiliary Lagrange multiplier
field. There is no electromagnetic vector potential in the action
since the field is electrically neutral in this
formulation. $\Gamma_{AB}$ refers not to the $U(1)$ phase symmetry but
to the two level symmetry. It plays the same essential role as
$a_\mu$.  Whereas $a_\mu$ was effectively off-diagonal in $\pm$
CTP-space, this is off-diagonal in the level space. This provides
another correspondence between the CTP and the two level system.

Variation of the action with respect to
$\Gamma(t)$ provides us with the conserved current.
\beq
\frac{\delta S}{\delta \Gamma_{AB}} = \frac{i}{2} (\psi_A^*\psi_B-\psi^*_B\psi_A)
\eeq
which represents the amplitude for stimulated transition between the
levels.  The current generated by this connection is conserved only on
average, since we are not taking into account any back-reaction. The
conservation law corresponds merely to
\beq
\partial_t \left(\frac{\delta S}{\delta \Gamma_{AB}} \right) \propto \sin(2\int X(t))
\eeq
where $X(t)$ will be defined later.
The potential $V(t)$ is time dependent and comprises the effect of the
level splitting as well as a perturbation mediated by the radiation
field. The `connection' $\Gamma_{21} = -\Gamma_{12}$ is introduced since
the diagonalization procedure requires a time-dependent unitary
transformation and thus general covariance demands that this will
transform in a different basis. The physics of the model depends on
the initial value of this `connection' and this is the key to the
trivial solubility of the Jaynes-Cummings model.

In matrix form we may write the action for the matter fields
\beq
S = \int dV_t \; \psi^*_A {\cal O}_{AB} \psi_B
\eeq
where
\beq
{\cal O}=\left[ 
\begin{array}{cc}
-\frac{\hbar^2\nabla^2}{2m}-V_1 -\frac{i\hbar}{2} D_t & J(t)+i\Gamma_{12}\\
J(t)-i\Gamma_{12} &-\frac{\hbar^2\nabla^2}{2m}-V_2 -\frac{i\hbar}{2}\stackrel{\leftrightarrow}{D_t}\\
\end{array}
\right].
\eeq
The level potentials may be regarded as constants in the effective
theory. They are given by
\beq
V_1 &=& E_1\nonumber\\
V_2 &=& E_2 - \hbar\Omega_R
\eeq
where $\hbar\Omega_R$ is the interaction energy imparted by the photon
during the transition i.e. the continuous radiation pressure on the
atom. In the effective theory we must add this by hand since we have
separated the levels into independent fields which are electrically
neutral, but it would follow automatically in a truly microscopic
theory in which all electromagnetic forces were included.  The quantum
content of this model is now that this recoil energy during a
transition is a quantized unit of $\hbar \Omega$, the energy of a
photon at the frequency of the source. Also the amplitude of the
source $J$ would be quantized and proportional to the number of
photons on the field.

To solve the problem posed in eqn. (\ref{eq1}), it is desirable to
perform a unitary transformation on the action
\beq
\psi &\rightarrow& U\psi\\
{\cal O} &\rightarrow& U{\cal O}U^{-1}
\eeq
which diagonalizes the operator $\cal O$.  The connection
 $\Gamma$ transforms under this procedure by
\beq
\Gamma \rightarrow \Gamma + \frac{i\hbar}{2} \left( U(\partial_t U^{-1}) - (\partial_t U)U^{-1} \right)
\eeq
since this requires a time-dependent transformation.
For notational simplicity we define $\hat L = -\frac{\hbar^2\nabla^2}{2m} -\frac{i}{2}\hbar\stackrel{\leftrightarrow}{D_t}$, so that the secular equation for the
action is:
\beq
(\hat L - E_1-\lambda(t))(\hat L - E_2 +\hbar\Omega -\lambda(t))\nonumber\\
 - (J^2(t)+\Gamma_{12}^2) = 0.
\eeq
Note that since $J\stackrel{\leftrightarrow}{\partial_t} J = 0$
there are no operator difficulties with this equation.
The eigenvalues are thus
\beq
\lambda_\pm &=& \hat L - \overline E_{12} + \hbar \Omega 
\pm \sqrt{\frac{1}{4}(\tilde E_{21}-\hbar\Omega)^2 + J^2(t)+\Gamma_{12}^2}\\
&\equiv& \hat L - \overline E_{12} + \hbar \Omega 
\pm \sqrt{\hbar^2\tilde\omega^2 + J^2(t)+\Gamma_{12}^2}\\
&\equiv&  \hat L - \overline E_{12} + \hbar \Omega \pm \hbar \omega_R,
\eeq
where $\overline E_{12} = \2(E_1+E_2)$ and $\tilde E_{21}=(E_2-E_1)$.
For notational simplicity we define $\tilde\omega$ and $\omega_R$ by
this relation. One may now confirm this procedure by looking for the
explicit eigenvectors and constructing $U^{-1}$ as the matrix of these
eigenvectors. This is done in ref. \cite{burgess15} and takes the form
\beq
U = \left( 
\begin{array}{cc}
\cos\theta & \sin\theta\\
-\sin\theta & \cos \theta
\end{array}
\right).
\eeq
In the diagonal basis, the centre of mass motion of the neutral atoms
factorizes from the wave-function, since a neutral atom in an
electromagnetic field is free on average. The two equations in the
matrix above are unravelled by writing the `gauge transformation'
\beq
\psi_\pm(x) = e^{\pm i \int_0^t X(t')dt'}\;\overline\psi(x),
\eeq
where $X(t)$ is presently undetermined and the wave-function for
the centre of mass motion in
$n=3$ dimensions,
\beq
\overline\psi(x) = \int \frac{d\omega}{(2\pi)}\frac{d^n{\bf k}}{(2\pi)^n}
\;e^{i({\bf k}\cdot{\bf x}-\omega t)}\delta \left(
\chi
\right)
\eeq
is a general linear combination of plane waves
satisfying the dispersion relation for centre of mass motion
\beq
\chi = \frac{\hbar^2{\bf k}^2}{2m} + \hbar(\Omega-\omega) - \overline E_{12} = 0.
\eeq
Substituting this form, we identify $X(t)$ as
the integrating factor for the uncoupled differential equations. The
complete solution is therefore
\beq
\psi_\pm(x) = e^{\mp i \int_0^t (\omega_R + i\partial_t\theta)dt'}\;\overline\psi(x).
\eeq
Notice that this result is an exact solution in the sense of being in
a closed form.  In the language of a gauge theory this result is gauge
dependent. This is because our original theory was not invariant under
time dependent transformations. The covariant procedure we have
applied is simply a method to transform the equations into an
appealing form; it does not imply invariance of the results under
a wide class of sources. On the other hand, it might be argued that
the invariant extension of eqn. (\ref{eq1}) should be considered.

The form of the action in eqn (\ref{eq1}) seems arbitrary and
unrelated to the earlier discussion but it may be placed in the
context of ref. \cite{burgess12} by an understanding of the conformal
nature of the perturbation. This includes an understanding of the
value of the connection $\Gamma_{12}$. The action is perturbed by a
time-dependent source which one hopes would lead to a stable theory
(the form of the action remaining the same over time). This suggests
precisely a measure of conformal covariance. In order to understand
this conformal connection in a familiar language, it is advantageous
to write the above model as the limit of a pseudo-relativistic theory
since the conformal group is an extension of the Lorentz group. This
also makes direct contact with ref. \cite{burgess12}.  The consistency
of such an approach has been verified in ref.
\cite{ourpaper}. Beginning with the Lorentz covariant action
\beq
S = \int dV_t \Big\{
\2 (\partial_\mu\phi_A)(\partial^\mu\phi_A) &+&\2 m_A\phi^2_A\nonumber\\
 &+& J_{AB}(t)\phi_A\phi_B
\Big\},
\eeq
where $J_{AB}$ now stands in place of $V_{AB}$ in the non-relativistic
formulation.
we consider a conformal rescaling by letting $g_{\mu\nu}\rightarrow
\Omega^2\;\overline g_{\mu\nu}$.  The action is not invariant under
this rescaling: if it were, there would be no need for the connection
$\Gamma$, or indeed this paper. The volume element scales as the
square root of the determinant of the metric,
i.e. $\sqrt{g}\rightarrow \Omega^4 \sqrt{\overline g}$ in $3+1$
dimensions, but we shall keep this separate for now. Since the issue
is not invariance but equivalence, this will not play a crucial
role. The first term in the braces contains one inverse power of the
metric, the second none and the third two. Choosing $\Omega^2 =
J_{AB}$, the off-diagonal, symmetric matrix with non-zero elements
$J(t)$, one can absorb the time dependent interaction by performing a
generalized rescaling. Rescaling the fields by
$\phi\rightarrow\Omega\overline\phi$, the action takes the form
\beq
S = \int dV_t' \Big\{
\2 (D_\mu\overline\phi_A)(D^\mu\overline\phi_A) &+&\2 m'_A\overline\phi^2_A \nonumber\\
&-&K^\mu_{AB} \overline
\phi(D_\mu\overline\phi)\Big\}
\eeq
where $D^\mu = \partial^\mu\delta_{AB} + K^\mu_{AB}$, which
is obtained by moving the scale factors through the derivatives, and
\beq
\2\frac{\partial_t \,J(t)_{AB}}{J(t)} = \left(\frac{\partial_t \Omega}{\Omega}\right)_{AB} = K_{AB}.
\eeq
$K$ is now analogous to $\Gamma$.
The familiar form of the conformal correction $\partial_\mu \Omega/\Omega$
is replaced in eqn. (\ref{eq1}) simply by $K_{12} = \partial_t \Omega$, which
makes the initial value of $K_{12}\sim \partial_t J$ clear: it
is the connection required for the derivatives to commute with a
conformal rescaling brought about by the perturbation.

The non-relativistic limit of the transformed action with a time-only
dependent $\Omega(t)$ leads to eqn. (\ref{eq1}), up to a Jacobian.
Thus although these actions are not identical, they are related by an
overall spacetime-dependent scale factor which behaves as though to
view the system through a distorting glass, exactly analogous to very
similar analysis of an effective non-equilibrium system in
ref. \cite{burgess10}. This is the price one pays for considering
partial systems.  The reason why these two theories give essentially
the same results is that they have the same structural elements. As
indicated in the introduction, it is this feature of effective field
theories which makes them robust and usable.

The field solutions to the two-level atom in the Jaynes-Cummings
approximation, near resonance, are known to exhibit so-called Rabi
oscillations, where the major populations oscillate between the upper
and lower levels. The oscillations in the general system are
dramatically different away from resonance, when one uses parameter
values which are appropriate for the micro-maser.  The outcome of this
analysis is perhaps surprising: the result which one obtains by making
the rotating wave approximation in a quantum mechanical formalism (the
Jaynes-Cummings model), coincidentally corresponding to a connection
$K_{12}\sim\sin(\Omega t)$, and this is a stable, steady state
effective field theory. The same action solved fully without
approximation $K_{12}$ corresponds to $K_{12}\sim 0$ and is not stable
to re-scalings; indeed it leads to extremely complicated
behaviour. That the stable model is identifiable with the
Jaynes-Cummings model is due as much to the method of analysis in time
dependent quantum mechanics as it is to do with the rotating wave
approximation.  The post Wilson renormalization group philosophy would
tend to favour a stable theory and say that another theory with
$K_{12}=0$ was not usefully predictable. Here we have an example where
the model can be solved, and indeed it does not appear to be give any
comparable pattern of behaviour to the Jaynes-Cummings model at
experimental values typical for the micro-maser, off
resonance\cite{jaynes1}, although it is clearly deterministic and
simple in structure.

\section{Conclusion}

The central message of this paper is that correct results in
non-equilibrium field theory are extremely sensitive to initial the
conditions and to the covariant structure of the action.  A gauge-like
formulation might allow us to solve inhomogeneous systems more easily,
but it also provides a conceptual perspective absent from the usual
diagrammatic approaches, showing us that approximation methods can
unwittingly change the effective structural elements of a theory and
produce completely different answers.  We make such approximations all
the time in interacting field theory.  Since the non-local sources
(which include the connection $a_\mu$) implement
`resummations'\cite{lawrie1,calzetta1,cjt} in interacting theories,
this must be of central importance in QCD and self-interacting models.

This work is supported by NATO
collaborative research grant CRG950018.  I am grateful to Meg
Carrington and Gabor Kunstatter for helpful discussions. Thanks
also to Cliff Burgess for comments.

\bibliographystyle{unsrt}

\end{document}